# Theory of THz generation by Optical Rectification using Tilted-Pulse-Fronts

Koustuban Ravi[1,*], Wenqian Ronny Huang[1], Sergio Carbajo[2,3], Emilio Nanni[1], Damian Schimpf[2], Erich.P.Ippen[1] and Franz.X.Kärtner[1,2,3]

*koust@mit.edu

[1] Department of Electrical Engineering and Computer Science, Research Laboratory of Electronics, Massachusetts Institute of Technology, Cambridge, MA 02139, USA
[2] Center for Free-Electron Laser Science, Deutsches Elektronen Synchrotron, Hamburg 22607, Germany
[3] Department of Physics and the Hamburg Center for Ultrafast Imaging, University of Hamburg, Germany

**Abstract:** A model for THz generation by optical rectification using tilted-pulse-fronts is developed. It simultaneously accounts for (i) the spatio-temporal distortions of the optical pump pulse, (ii) the nonlinear coupled interaction of THz and optical radiation in two spatial dimensions (2-D), (iii) self-phase modulation and (iv) stimulated Raman scattering. The model is validated by quantitative agreement with experiments and analytic calculations. We show that the optical pump beam is significantly broadened in the transverse-momentum ($k_x$) domain as a consequence of the spectral broadening caused by THz generation. In the presence of this large frequency and transverse-momentum (or angular) spread, group velocity dispersion causes a spatio-temporal break-up of the optical pump pulse which inhibits further THz generation. The implications of these effects on energy scaling and optimization of optical-to-THz conversion efficiency are discussed. This suggests the use of optical pump pulses with elliptical beam profiles for large optical pump energies. It is seen that optimization of the setup is highly dependent on optical pump conditions. Trade-offs of optimizing the optical-to-THz conversion efficiency on the spatial and spectral properties of THz radiation is discussed to guide the development of such sources.

## 1. Introduction

Terahertz (THz) sources are characterized by wavelengths, roughly hundred times larger than optical and ten times smaller than radio frequency sources. These wavelengths make THz radiation with high peak fields uniquely amenable to non-linear spectroscopy [1]-[2], compact particle-acceleration [3-6],high-harmonic generation [7], molecular alignment [8] etc .

Of various high field THz generation modalities, optical rectification (OR) of femtosecond laser pulses with tilted-pulse-fronts in lithium niobate has emerged as the most efficient THz generation technique. Among various efforts, it was developed in [9-13] as a means to achieve phase-matching in materials with large disparities between THz and optical refractive indices.

In this approach, an optical pump pulse is angularly dispersed to produce an intensity front which is tilted with respect to its propagation direction. THz radiation propagating perpendicular to this tilted intensity front or tilted-pulse-front (TPF) is then generated. Since the optical and THz radiation travel different distances in the same time, the difference between optical and THz refractive indices is compensated and phase-matching is achieved. OR using TPF's has resulted in optical-to-THz conversion efficiencies (henceforth referred to as conversion efficiency) in excess of 1% [14-15] and the highest THz pulse energy of 0.4 mJ [16] to date. Therefore, the approach is promising for the development of laboratory scale THz sources with pulse energies greater than mJ level. Comprehensive theoretical models to aid understanding and quantitatively predict the performance of such systems are therefore of interest. The requisites of a physically accurate model and the current state of theory are described below.

As a consequence of the angular dispersion of the optical pump pulse in OR using TPF's, various frequency components of the optical pump pulse spectrum are spatially separated. This is tantamount to having different spectral bandwidths, pulse durations and average frequency at each spatial location. These effects are termed spatio-temporal distortions [17] and affect the properties of the generated THz radiation. Secondly, since the generated THz propagates perpendicular to the TPF, the optical pump and THz radiation propagate non-collinearly. Most importantly, as THz radiation is generated, it is accompanied by a dramatic cascaded frequency down-shift and spectral broadening of the optical pump pulse spectrum (cascading effects). On one hand cascading is responsible for conversion efficiencies which exceed the Manley-Rowe limit. On the other hand, in the presence of group velocity dispersion due to angular dispersion (GVD-AD) and material dispersion (GVD-MD), this spectral broadening inhibits further THz generation [18-20]. A comprehensive theoretical model should therefore be able to account for all of the above effects. This would require a simultaneous solution of optical and THz electric fields (henceforth referred to as field) in at least two spatial dimensions (2-D). In addition, spatio-temporal distortions imparted by the TPF setup would also have to be considered.

Previously presented models broadly comprise of (i) 1-D and 2-D spatial models without the inclusion of cascading effects (i.e., nonlinear coupling between THz and optical radiation is not considered) and (ii) 1-D spatial models which account for cascading effects. Of works in category (i), a one-dimensional (1-D) spatial model including the effects of material dispersion and GVD-AD was presented in [21-22]. In [23], a 1-D model considering material dispersion, GVD-AD and self-phase modulation (SPM) was presented. In [24], a 2-D model which took into account material dispersion, GVD-AD and crystal geometry was developed. In category (ii), an effective 1-D model with cascading effects was first presented in [25] and improved in [18].

In this paper, we present the formulation of a 2-D model which simultaneously accounts for spatio-temporal distortions of the optical pump pulse, cascading effects, SPM and stimulated Raman scattering (SRS), material dispersion, THz absorption as well as geometry of the nonlinear crystal. The developed model is applicable to the simulation of a variety of OR systems with different TPF setups, crystal geometries and optical pump pulse formats.

In Section 2, we outline our general approach. In Section 3, we describe the physics of THz generation using OR with TPF's. In particular, a discussion from transverse momentum ($k_x$) and time domain viewpoints is introduced. It is seen that the generation of THz results in the broadening of the optical pump pulse in both frequency and transverse-momentum domains. In the presence of this increased frequency and transverse-momentum spread, GVD-AD and GVD-MD cause a spatio-temporal break-up of the optical pump pulse which inhibits further THz generation. These descriptions serve to motivate the theoretical formulation, which is presented in Section 4. In Section 5, we validate the model by comparisons to experiments and analytic calculations. The impact of imaging errors on conversion efficiency is shown quantitatively. It is seen that small perturbations to the optimal imaging configuration can result in sizeable degradation of conversion efficiency. Insights into the experimentally measured broadening of the optical spectrum are provided. In Section 6, we discuss the meaning of effective propagation length in 2-D. This is then used to discuss scaling to large optical pump energies and optimization of conversion efficiency. It is seen that the optimization of conversion-efficiency is highly dependent on the optical pump parameters. Finally, we highlight the trade-offs incurred while optimizing the conversion efficiency on spatial and spectral properties of THz radiation. We conclude in Section 7. This paper thus provides an overview for constructing sources customized optimally for various applications.

## 2. General Approach

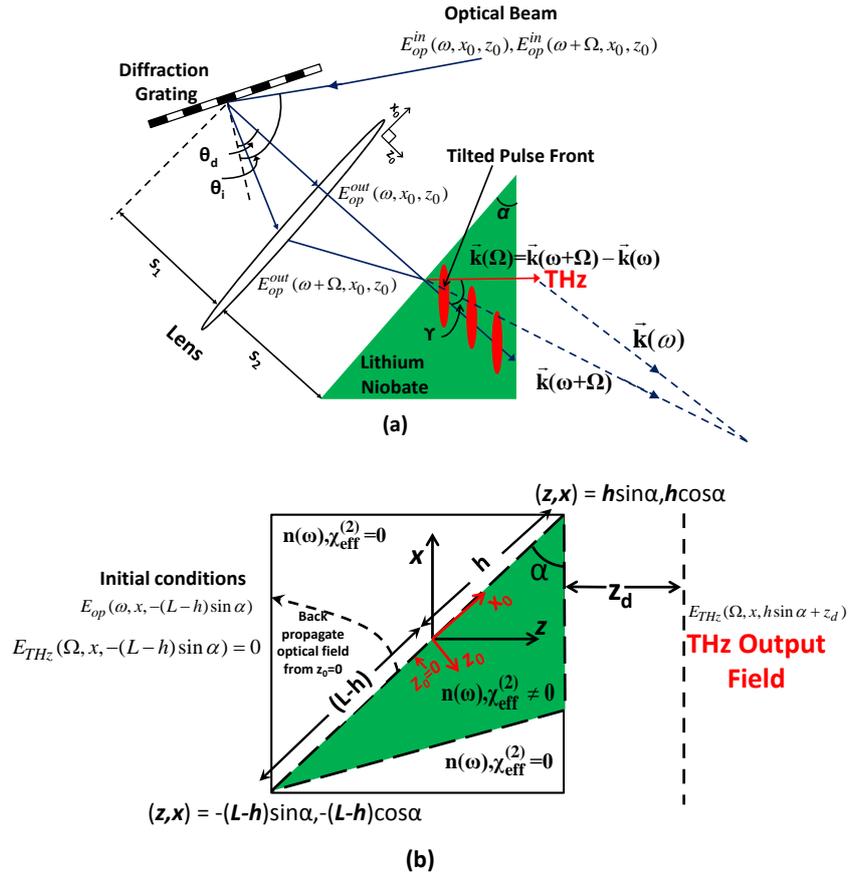

Fig. 1: Schematic of approach. (a) An optical pump pulse with electric field $E_{op}^{in}(\omega, x_0, z_0)$ is incident on a setup to generate a tilted-pulse-front. The model accounts for the angular dispersion of various spectral components which can generate THz radiation inside the nonlinear crystal by satisfying the appropriate phase-matching condition for optical rectification. From a time-domain viewpoint, the angularly dispersed pulse forms a tilted-pulse-front shown by the red ellipses. THz radiation is generated perpendicular to this tilted-pulse-front (red arrow). (b) 2-D computational space for solving coupled nonlinear wave equations for optical rectification. Nonlinear crystal geometry is accounted for by delineating an appropriate distribution of $\chi_{eff}^{(2)}(x,z)$. Edges of the distribution are smoothed out to avoid discontinuities. The refractive index is homogeneously distributed throughout the computational space. The optical beam is centered at a distance *h* from the apex of the crystal which sets the limits to the computational region. The optical field at the beginning of the lattice is calculated analytically using dispersive ray pulse matrices. The THz field profile can be calculated at a distance $z_d$ from the crystal after Fresnel reflection is taken into account.

The overall schematic of our approach is depicted in Fig. 1. An optical pump pulse with input electric field described by the complex variable $E_{op}^{in}(\omega, x_0, z_0)$ at angular frequency *ω*, propagating in the $z_0$ direction is incident on a TPF setup. In Fig. 1, a commonly employed setup incorporating a diffraction grating and single lens is depicted. However, the model is applicable to a variety of TPF setups (e.g. telescope and diffraction grating, contact grating etc.). In Fig. 1(a), electric fields at two angular frequencies *ω* and *ω+Ω* are depicted for convenience although there are infinitely many frequency components.

An optical pulse with a TPF is typically angularly dispersed [17]. Therefore, various frequency components of the emergent optical field described by $E_{op}^{out}(\omega, x_0, z_0)$ propagate at different angles. This is depicted in Fig. 1(a) as spectral components at *ω* and *ω+Ω* emerge with wave vectors (henceforth referred to as momentum) $\vec{k}(\omega), \vec{k}(\omega+\Omega)$ respectively. In the time domain, such an angularly dispersed pulse has an intensity profile tilted with respect to its propagation direction as shown by the red ellipses in Fig. 1(a). This gives rise to the terminology of 'tilted-pulse-front'. These red ellipses make an angle *π/2-γ* with respect to the propagation direction of the optical pulse, where *γ is* termed the pulse-front-tilt angle. Since OR is intra-pulse difference frequency generation (DFG), in the phase-matched condition, the momentum of the generated THz (at angular frequency *Ω*) is given by $\vec{k}(\Omega) = \vec{k}(\omega+\Omega) - \vec{k}(\omega)$ as depicted by the red arrow in Fig. 1(a). The generated THz then emerges at an angle *γ* with respect to the direction of propagation of the optical pump pulse and exits approximately normal to the output facet. Since the various frequency components of the optical pump pulse are angularly separated, there is a spatial variation in the average frequency (spatial-chirp) as well as amplitude across the optical beam profile. As a consequence, the pulse durations and bandwidths at various points in space are different, which affects the spatial and spectral properties of the generated THz pulse. In our model, we consider these effects by applying an analytic formulation of dispersive ray pulse matrices from [26] in Section 4.1. This approach efficiently models the various spatio-temporal distortions associated with the optical pulse for an arbitrary TPF setup and supplies $E_{op}^{out}(\omega, x_0, z_0 = 0)$ at the input face of the nonlinear crystal as shown in Fig. 1(b).

The incident optical field excites a polarization in the nonlinear material to drive the generation of THz radiation. The generated THz in turn influences the propagation of the optical field and vice versa. Thus, the evolution of the optical and THz fields is described by a solution of a system of 2-D coupled nonlinear wave equations in the (*z-x*) co-ordinate system depicted in Fig. 1(b). In our approach, the (*z-x*) co-ordinate system is rotated with respect to ($z_0$-$x_0$) by an angle *α*, which is the apex angle of the nonlinear crystal. The angle $\alpha$ is approximately equal to the pulse-front-tilt angle $\gamma$. The rotated co-ordinate system then has two key advantages. Firstly, in this set of axes, the THz radiation has small transverse-momentum components, i.e. $k_x \sim 0$, which relaxes the constraints on spatial resolution $\Delta x$ and consequently alleviates computational cost. Secondly, it makes it convenient to include the transmission of THz radiation at the crystal boundary.

In Fig. 1(b), we delineate how we consider the geometry of the crystal. An extended Cartesian space in the ($z$-$x$) co-ordinate system, uniformly filled with material of refractive index $n(\omega)$ is considered. Only regions of the computational space physically occupied by the crystal would have a non-zero value of second order susceptibility $\chi^{(2)}_{eff}(x,z)$ as shown in Fig. 1(b). If the length of the input crystal face is $L$ and the optical field is incident at a distance $h$ from the apex, the computational region extends from $\left(-(L-h)\sin\alpha, -(L-h)\cos\alpha\right)$ to $\left(h\sin\alpha, h\cos\alpha\right)$ as shown in Fig. 1(b). The initial optical field profile along the line $z = -(L-h)\sin\alpha$ can be calculated analytically by back-propagating the optical field calculated at the input crystal face at $z_0$=0. In this model, we assume that the THz field is zero at the beginning of the computational space and consider only a single passage of the optical and THz beams through the crystal. In the limit of relatively thick crystals where the reflected THz energy is absorbed (absorption length at 300K is~ 2 mm) or with the use of THz anti-reflection coatings, this approximation is well justified.

## 3. Physical description: spatio-temporal break-up of the optical pump pulse

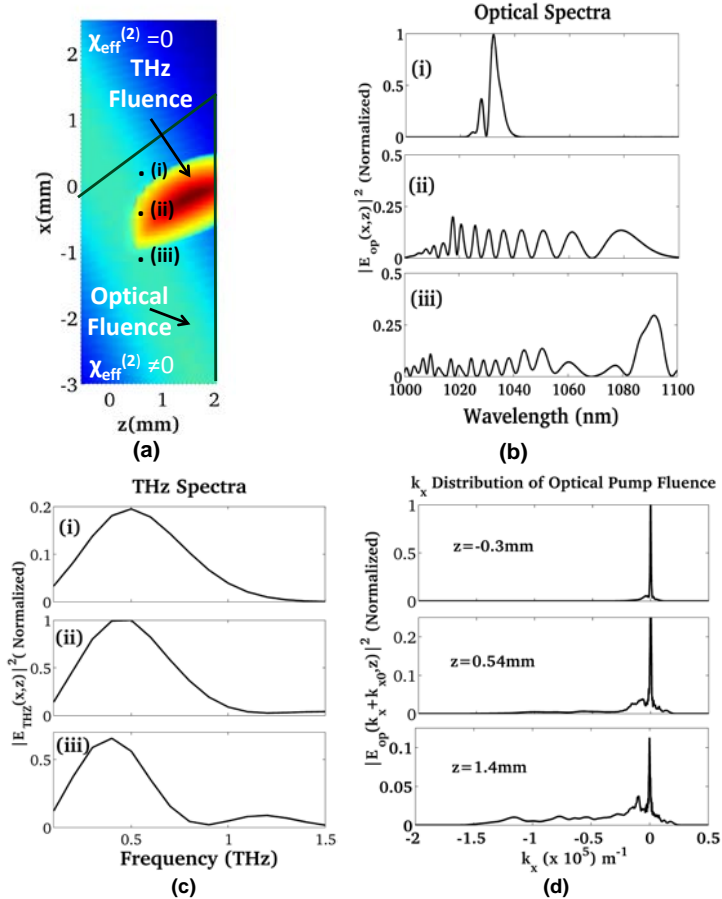

Fig. 2(a). Spatial distribution of the optical and THz fluences: The THz field propagates in the *z* direction with the optical field at an angle *γ*~63° with respect to it. THz is only generated over a small portion of the optical field due to spectral broadening of the optical pulse by cascading which results in disruption of phase-matching due to group velocity dispersion (by angular and material dispersion) for subsequent portions of the beam. (b) Optical spectrum is broadened between (i)-(iii) due to cascading effects. (c) THz spectra at locations (i)-(iii) show significant spatial-chirp due to spatial variations of the optical electric field in (c). (d) Since each frequency component has a certain value of transverse-momentum in an angularly dispersed beam, spectral broadening also necessarily results in broadening in transverse-momentum $k_x$. As the optical spectrum broadens, there is a broadening in transverse-momentum between z = -0.3 mm and 1.4 mm.

In Fig. 2, we provide a sample solution using the developed model. This will bring out the essential physics and put the formulation in Section 4 into context. Figure 2(a), depicts the optical and THz fluences as a function of space. The region within the green lines has non-zero $\chi^{(2)}_{eff}(x,z)$. The optical pump pulse propagates obliquely in the nonlinear crystal as indicated by the cyan/light-blue colormap in Fig. 2(a). The THz pulse however propagates in the *z* direction, at an angle *γ*~63° to the optical pump pulse and emerges perpendicular to the output face as shown by the THz fluence in the red colormap.

As the optical pump pulse propagates, it generates THz photons at $\Omega$ and simultaneously suffers a frequency downshift by the same amount. With successive generation of THz photons it repeatedly experiences a frequency down-shift or 'cascaded' frequency down-shift which leads to large spectral broadening of the optical pump pulse spectrum as seen in Fig. 2(b). Even if the total depletion of the optical pump energy is only 1%, the drastic spectral reshaping renders undepleted pump approximations inaccurate.

As the optical pump pulse spectrum is modified between locations (i)-(iii), the subsequent THz spectrum is also modified as shown in Fig. 2(c) and vice-versa. As a result, there is significant spatial variation in both optical and THz spectra. The generated THz spectra are broadband, extending from 0 to 1 THz, consistent with earlier experiments and theory.

For an angularly dispersed pulse, each spectral component of the optical pulse at *ω* has a well-defined transverse-momentum $k_x$ (smaller *ω*'s have a more negative $k_x$ value as shown in Fig. 1(a)). Therefore, spectral broadening of the optical pulse also directly leads to re-distribution of optical pulse energy among various transverse momentum values $k_x$ as seen in Fig. 2(d). The spread in transverse momentum in Fig. 2(d) is on the order of $10^4 \text{m}^{-1}$, which is still much smaller than the optical wave number (~$10^6 \text{m}^{-1}$) which means the beam is still relatively paraxial.

In the presence of this large spectral broadening one would expect a temporal break-up of the pulse due to group velocity dispersion. In addition, due to broadening in transverse momentum, there is also a spatial break-up of pulses. In combination, a rapid spatio-temporal break-up of the optical pulse occurs as shown in Fig. 3. Thus an initially clean TPF in Fig. 2(a) suffers a spatio-temporal break-up upon propagation in Fig. 2(c) as it propagates over very short distances on the order of ~ 2 mm. Due to this spatio-temporal break-up, different parts of the optical pulse arrive at different times and the generated THz no longer builds up coherently.

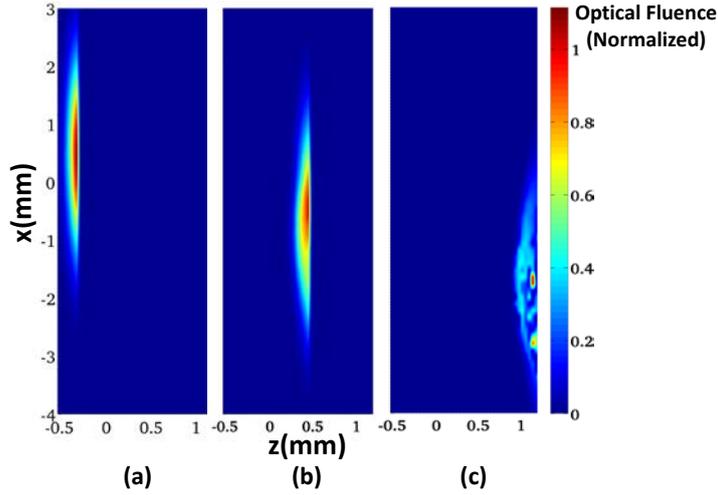

Fig. 3: Spatio-temporal break-up of the optical pulses due to simultaneous broadening in frequency and transverse momentum $k_x$. (a) The tilted pulse-front is parallel to the output facet. (b) Due to group-velocity dispersion effects (dominated by GVD-AD which is ~15 times larger than GVD-MD), the pulse is broadened in time and space. (c) A spatio-temporal break-up of the pulse occurs and different parts of the optical pump arrive at different times, preventing further coherent build-up of THz radiation.

## 4. Theory

*4.1 Propagation of Optical Pump Fields through the Optical Setup*

As described in Section 2 and depicted in Fig. 1, a TPF setup imparts a number of spatio-temporal distortions to the optical pump pulse which influences the properties of the generated THz radiation.

In this section, we show how to account for these effects for an arbitrary TPF setup by employing dispersive ray pulse matrices [26]. Although developed for passive optical elements, our application of this approach to OR using TPF's results in a powerful model, closely connected to experiments. An explicit expression for the electric field of the optical pump pulse is obtained which allows calculations to be performed rapidly. Note that alternate ray-pulse matrix approaches such as [27] are also applicable. Since the beam size of the optical pump used in OR is much larger than the optical wavelength, paraxial approximations of ray-pulse matrix schemes are valid for the optical pump. Each spectral component $E_{op}^{in}(\omega, x_o, z_o)$ of the optical pump pulse at angular frequency $\omega$, with input beam centred at a transverse position $x_{in}(\omega)$ and propagation direction $x'_{in}(\omega)$ emerges with transverse position $x_{out}(\omega)$ and propagation direction $x'_{out}(\omega)$ from the optical setup, just before entering the nonlinear crystal.

The dependence of the position and propagation direction on frequency accounts for spatial-chirp and angular dispersion. The propagation direction here is not the physical direction but one normalized by the refractive index of the medium [26] which makes it convenient to apply to systems with interfaces of mismatched refractive index. The relationship between $x_{in}(\omega)$, $x'_{in}(\omega)$ and $x_{out}(\omega)$, $x'_{out}(\omega)$ is given by Eq. (1). Here, $\overline{\overline{M(\omega)}}$ is the overall ray-pulse matrix obtained by the product of ray-pulse matrices of individual optical components $\overline{\overline{M_i(\omega)}}$ in reverse order of incidence. The beam positions and propagation directions after the $i^{th}$ optical element are $x_{out_i}(\omega)$ and $x'_{out_i}(\omega)$.

$$\overline{\overline{M(\omega)}} \begin{bmatrix} x_{in}(\omega) \\ x'_{in}(\omega) \\ 1 \end{bmatrix} = \begin{bmatrix} x_{out}(\omega, z_0 = 0) \\ x'_{out}(\omega) \\ 1 \end{bmatrix} \qquad (1)$$

The ray pulse matrix of the $i^{th}$ optical component $\overline{\overline{M_i(\omega)}}$ is described by a 3x3 matrix for a single transverse spatial dimension as shown in Eq. (2).

$$\overline{\overline{M_i(\omega)}} = \begin{bmatrix} \left.\frac{\partial x_{out_j}}{\partial x_{in}}\right|_i & \left.\frac{\partial x_{out}}{\partial x'_{in}}\right|_i & \left.\frac{\partial x_{out}}{\partial \omega}\right|_i (\omega - \omega_0) \\ \left.\frac{\partial x'_{out}}{\partial x_{in}}\right|_i & \left.\frac{\partial x'_{out_j}}{\partial x'_{in}}\right|_i & \left.\frac{\partial x'_{out}}{\partial \omega}\right|_i (\omega - \omega_0) \\ 0 & 0 & 1 \end{bmatrix} = \begin{bmatrix} A_i(\omega) & B_i(\omega) & E_i(\omega) \\ C_i(\omega) & D_i(\omega) & F_i(\omega) \\ 0 & 0 & 1 \end{bmatrix} \qquad (2)$$

Equation (2) shows that the upper 2x2 matrix is nothing but the standard ABCD matrix for Gaussian beams. However, in order to account for dispersion, there are two additional terms $E_i$ and $F_i$ which correspond to the partial derivatives $\left.\partial x_{out}/\partial \omega\right|_i (\omega - \omega_0)$ and $\left.\partial x_{out}'/\partial \omega\right|_i (\omega - \omega_0)$. These terms refer to the shift in output beam position and output beam propagation direction in response to a shift in frequency. Here, we calculate $F_i$ upto the fourth order in frequency and accounts for GVD-AD and higher order terms. Note that the last row of $\overline{\overline{M_i(\omega)}}$ is [0 0 1] as the source frequency does not change. With the knowledge of the input and output beam positions and propagation directions, a Huygen's integral can be used [26] to calculate the electric field of the emergent optical pump pulse after it has passed through the TPF setup as shown in Eq.(3).

$$E_{op}^{out}(\omega, x_0, z_0) = \sqrt{\frac{w_{in}(\omega)}{w_{out}(\omega, z_0)}} \cdot E_0(\omega) e^{-jk(\omega)\frac{(x_0 - x_{out}(\omega, z_0))^2}{2q_{out}(\omega, z_0)}} e^{-jk(\omega)z_0} e^{-jk(\omega)x'_{out}(\omega)x_0} e^{-j\phi(\omega, z_0)}$$

3(a)

$$\sqrt{\frac{1}{A(\omega, z_0) + B(\omega, z_0)/q(\omega)_{in}}} = \sqrt{\frac{w_{in}(\omega)}{w_{out}(\omega, z_0)}} e^{-j\varphi_1(\omega, z_0)} \qquad 3(b)$$

$$\phi_2(\omega, z_0) = \frac{k(\omega)}{2}[x_{in}(\omega, z_0)x'_{in}(\omega) - x_{out}(\omega, z_0)x'_{out}(\omega)] \qquad 3(c)$$

$$\phi_3(\omega, z_0) = \frac{k(\omega)}{2} \sum_{i=1}^{N} F_i(\omega) x_{out_{i+1}}(\omega) \qquad 3(d)$$

In Eq. 3(a), $w_{in}(\omega), w_{out}(\omega, z_0)$ are the $e^{-2}$ beam radii of the spectral component at $\omega$ at the input and output of the TPF setup respectively. The pre-factor $\sqrt{w_{in}(\omega)/w_{out}(\omega, z_0)}$ in Eq. 3(a) represents the change in the optical field intensity due to change in beam size. $E_0(\omega)$ represents the initial spectral amplitude of the electric field at angular frequency $\omega$ and may include phase terms describing chirp. The first exponential term in Eq. 3 (a), represents the transverse variation of the beam along $x_0$, where $q_{out}$ is the usual $q$ parameter associated with Gaussian beams. Notice that each spectral component is centered at a different beam position $x_{out}(\omega, z_0)$, which accounts for spatial-chirp. The second exponential term $e^{-jk(\omega)z_0}$ is the phase associated with the roughly $z_0$ propagating paraxial field. The wave number $k(\omega) = \omega n(\omega)/c$ includes the material dispersion of the medium at optical frequencies. The third exponential term $e^{-jk(\omega)x'_{out}(\omega)x_0}$ represents angular dispersion which directly influences phase-matching. Since the optical pump beam is treated paraxially, $x'_{out}(\omega) \ll 1$. Since, the value of $x'_{out}(\omega)$ is mapped to the configuration of the TPF setup (grating angle, imaging distances etc.), Eq. 3(a) can quantify the impact of imaging errors on the properties of THz radiation (spectrum, spatial profile, conversion efficiency). The various phase terms in Eq. 3(a), $\phi(\omega) = \phi_1(\omega) + \phi_2(\omega) + \phi_3(\omega)$ are given by Eqs. 3(b)-3(d). The $\phi_1(\omega)$ term represents the phase correction due to a Gaussian beam. The remaining terms $\phi_2(\omega)$ and $\phi_3(\omega)$ account for additional phases introduced by the TPF setup. Thus using these dispersive ray-pulse matrices, one obtains an explicit expression for the electric $E_{op}^{out}(\omega, x_0, z_0)$ field for every spectral component of the optical pump pulse spectrum. This calculated $E_{op}^{out}(\omega, x_0, z_0)$ value is used as an initial condition to solve the nonlinear coupled system of wave equations. This expression accounts for spatial variations, material dispersion, angular dispersion including GVD-AD, spatial frequency-variations and spatial variations in pulsewidth. It allows for a one-to-one correspondence between the TPF setup configuration and the properties of THz radiation to be established.

*4.2 Nonlinear Polarization due to Optical Rectification*

In Section 4.1, we obtained the electric field of the optical pump pulse inside the crystal in the co-ordinate system ($z_0$-$x_0$) as shown in Fig.1. In this section, we calculate the nonlinear polarization terms which drive the optical and THz fields. The nonlinear polarization will be calculated in the ($z$-$x$) co-ordinate system introduced in Fig.1 (b). The transformation between ($z_0$-$x_0$) and ($z$-$x$) co-ordinate systems is easily obtained via Eq. (4).

$$x_0 = x\cos\alpha + z\sin\alpha, \quad z_0 = -x\sin\alpha + z\cos\alpha \qquad (4)$$

We use Eq. (4) in Eq. (3), to obtain the electric field of the optical pump pulse in the transformed co-ordinates i.e. $E_{op}^{out}(\omega, x, z)$. The value of $E_{op}^{out}(\omega, x, z)$ at $z = -(L-h)\sin\alpha$ serves as the initial condition for the evolution of wave equations described in Section 4.3.

Equation (5) describes the nonlinear polarization term $P_{THz}(\Omega, x, z)$ due to OR which drives the THz electric field $E_{THz}(\Omega, x, z)$ at the angular frequency $\Omega$.

$$P_{THz}(\Omega, x, z) = \varepsilon_0 \chi_{eff}^{(2)}(x, z) \cdot \int_0^\infty E_{op}(\omega + \Omega, x, z) E_{op}^*(\omega, x, z) d\omega \qquad (5)$$

In Eq. (5), $E_{op}(\omega + \Omega, x, z)$ corresponds to the electric field of the spectral component of the optical pump pulse spectrum at angular frequency $\omega$ and spatial location ($z,x$). The nonlinear polarization term at each spatial location ($z,x$) in Eq. (5) can be seen to be an aggregate of all possible DFG processes between the spectral component $E_{op}(\omega+\Omega,x,z)$ and $E_{op}(\omega,x,z)$. In Eq. (5), $\chi_{eff}^{(2)}(x,z)$ is the effective second order nonlinear susceptibility for OR at each spatial location and $\varepsilon_0$ is the free space permittivity. The spatial dependence of the effective non-linear susceptibility is used to account for the geometry of the nonlinear crystal as was shown in Fig. 1(b).

If one substitutes the expression for the electric field from Eq. 3(a) in Eq.(5), various factors which affect phase-matching effects become evident. For instance, because of the angular dispersion term in Eq. 3(a), a term proportional to $\exp\left\{-j\left[k(\omega+\Omega)x'_{out}(\omega+\Omega)-k(\omega)x'_{out}(\omega)\right]x_0\right\}$ appears in Eq. (5) which enables us to quantify the effect of phase-mismatch due to imaging errors. Similarly, a term proportional to $\exp\left\{-j\left[\frac{\omega+\Omega}{c}\frac{(x_0-x_{out}(\omega+\Omega))^2}{2q_{out}(\omega+\Omega)}-\frac{\omega}{c}\frac{(x_0-x_{out}(\omega))^2}{2q_{out}(\omega)}\right]\right\}$ describes a spatial variation in the magnitude of $P_{THz}(\Omega,x,z)$ leading to spatial variation of the generated THz frequency. This term also describes the effects of the finite radius of curvature of the optical pump pulse which affects phase-matching. In addition, there is a term of the form $\exp\left\{-j[\phi(\omega+\Omega)-\varphi(\omega)]\right\}$ which introduces phase mismatch due to various phase accumulations through the optical setup.

Similarly, each spectral component $E_{op}(\omega,x,z)$ of the optical pump pulse spectrum is driven by a nonlinear polarization term $P_{op}(\omega,x,z)$ described in Eq. (6).

$$\begin{aligned}P_{op}(\omega,x,z) = & \varepsilon_0\chi_{eff}^{(2)}(x,z).\int_0^\infty E_{op}(\omega+\Omega,x,z)E_{THz}*(\Omega,x,z)d\Omega \\ & +\varepsilon_0\chi_{eff}^{(2)}(x,z).\int_0^\infty E_{op}(\omega-\Omega,x,z)E_{THz}(\Omega,x,z)d\Omega \\ & -\frac{2k_{z0}(\omega)\varepsilon_0 c^2}{\omega^2}\mathbf{F}_t\left\{\frac{\varepsilon_0\omega_0 n(\omega_0)^2 n_2(x,z)}{2}|E_{op}(t,x,z)|^2 E_{op}(t,x,z)\right\} \\ & -\frac{2k_{z0}(\omega)\varepsilon_0 c^2}{\omega^2}\mathbf{F}_t\left\{j\frac{\varepsilon_0\omega_0 n(\omega_0)^2 n_2(x,z)}{2}\left[|E_{op}(t-t',x,z)|^2\otimes h_r(t')\right]E_{op}(t,x,z)\right\}\end{aligned} \quad (6)$$

The first term in Eq. (6) is the analogue term on the right hand side of Eq. (5). It signifies that an optical photon at angular frequency $\omega$ is created by an aggregate of DFG processes between optical photons at angular frequency $\omega+\Omega$ and THz photons at angular frequency $\Omega$. It represents the red-shift of the optical spectrum depicted in Fig. 2(b). The second term corresponds to an aggregate of SFG processes between optical photons at angular frequency $\omega-\Omega$ and THz photons at angular frequency $\Omega$. This term partially contributes to the blue-shift of the optical spectrum that was seen in Fig. 2(b). The third term in Eq. (6) represents the SPM term. Here, $E_{op}(t,x,z)$ is the time-domain electric field of the optical pump pulse and $\mathbf{F}_t$ represents the Fourier transform between time and frequency domains. The intensity dependent refractive index coefficient is given by $n_2(x,z)$. Since the SPM term in Eq. (6) contains details of the spatial distribution of the optical field, it also accounts for self-focusing effects. The final term models Stimulated Raman Scattering. This term is related to the SPM term but includes the effects of a Raman gain lineshape given by $h_R(\omega')$.

*4.3 Solving the 2-D non-linear wave equation using Fourier Decomposition*

In this section, we present our approach for solving the coupled system of nonlinear wave equations. The nonlinear polarization terms defined in Eqs. (5) and (6) will drive the corresponding THz and optical fields. The nonlinear scalar wave equation for the evolution of the THz field $E_{THz}(\Omega, x, z)$ is presented in Eq. (7). A single scalar wave equation, treating one vector component of the THz beam would suffice for lithium niobate since the $d_{33}$ element of the second order nonlinear tensor is much larger than other $d_{ij}$ and THz generation scales as $d_{ij}^2$.

$$\nabla^2 E_{THz}(\Omega, x, z) + k^2(\Omega) E_{THz}(\Omega, x, z) = \frac{-\Omega^2}{\varepsilon_0 c^2} P_{THz}(\Omega, x, z) \qquad (7)$$

In Eq. (7), $k(\Omega) = \Omega n(\Omega)/c$ is the wave number at the THz angular frequency $\Omega$ and $n(\Omega)$ is the corresponding refractive index. Similar to Eq. (7), one can also write the corresponding wave equation for the optical fields at various angular frequencies $\omega$ in Eq. 8(a).

$$\nabla^2 E_{op}(\omega, x, z) + k^2(\omega) E_{op}(\omega, x, z) = \frac{-\omega^2}{\varepsilon_0 c^2} P_{op}(\omega, x, z) \qquad 8(a)$$

In the (*z-x*) co-ordinate system, the optical radiation is propagating at a large angle of ~ 63° relative to the THz radiation. Furthermore, the optical wave number is a hundred times larger than the THz wave number. Therefore, the spread in transverse momentum required to account for the oblique propagation of the optical field would be very large. Direct solutions of Eqs. (7) and 8(a) will therefore be cumbersome because of the fine spatial resolution that will be required to account for this large spread in transverse momentum. In order to provide an efficient solution to the problem, we define a solution of the form $E_{op}(\omega, x, z) = A_{op}(\omega, x, z) e^{jk_{x0}(\omega).x} e^{-jk_{z0}(\omega).z}$. Here, $k_{x0}(\omega)$ and $k_{z0}(\omega)$ are the momentum components in the *x* and *z* directions and are functions of $\omega$ due to angular dispersion. This expression signifies that the optical beam has a narrow spread in transverse-momentum in comparison to $k_{x0}(\omega)$. Equation 8 (a) then reduces to Eq. 8(b) for the evolution of the optical field. The second derivative in *z* is dropped since $|k_{z0} A'_{op}(\omega, z)| \gg |A_{op}''(\omega, z)|$. The second derivative in *x* is retained to account for the radius of curvature of the Gaussian beam.

$$-2jk_{z0}(\omega) \frac{\partial A_{op}(\omega, x, z)}{\partial z} + 2jk_{x0}(\omega) \frac{\partial A_{op}(\omega, x, z)}{\partial x} + \frac{\partial^2 A_{op}(\omega, x, z)}{\partial x^2}$$
$$= \frac{-\omega^2}{\varepsilon_0 c^2} P_{op}(\omega, x, z) e^{-jk_{x0}(\omega)x + jk_{z0}(\omega)z} \qquad 8(b)$$

Eq. (7) drives the THz field which in turn affects the optical field via the nonlinear polarization term $P_{op}(\omega, x, z)$ defined in Eq. (6). However, the optical field directly influences the THz field via the nonlinear polarization term $P_{THz}(\Omega, x, z)$ defined in Eq. (5). Thus, Eqs. (7) and 8(a) form a coupled system of wave equations for THz and optical fields. An elegant solution to this system can be obtained via spatial Fourier decomposition. This effectively breaks up Eqs. (7) and 8(b) into a system of coupled 1-D first order differential equations which can be solved highly efficiently as will be shown below.

Applying Fourier transforms to both the left and right hand sides of Eq. (7), we obtain Eq. (9a), which is a 1-D differential equation in $z$ for $E_{THz}(\Omega, k_x, z)$. The evolution of $E_{THz}(\Omega, k_x, z)$ at each $\Omega$ and $k_x$ can be updated in parallel.

$$\frac{\partial^2 E_{THz}(\Omega, k_x, z)}{\partial z^2} + \left(k^2(\Omega) - k_x^2\right) E_{THz}(\Omega, k_x, z) = \frac{-\Omega^2}{\varepsilon_0 c^2} P_{THz}(\Omega, k_x, z) \qquad 9(a)$$

Equation 9(a) can be further simplified to Eq. (9b) by assuming a solution of the form $E(\Omega, k_x, z) = A(\Omega, k_x, z)e^{-jk_z(\Omega, k_x)z}$ which is a slow varying envelope approximation, where $k_z(\Omega, k_x) = \sqrt{k^2(\Omega) - k_x^2}$. This is not to be confused for a paraxial approximation for the THz field since each component has a different $k_z(\Omega, k_x)$, which allows for deviations of $\pm \pi/2$ from the $z$ direction. However, back-propagating THz components are ignored as they would not be well phase matched. The THz absorption coefficient at angular frequency $\Omega$ is given by $\alpha(\Omega)$.

$$\frac{\partial A(\Omega, k_x, z)}{\partial z} = \frac{-\alpha(\Omega)}{2} A(\Omega, k_x, z) - \frac{j\Omega^2}{(2k_z(\Omega, k_x)\varepsilon_0 c^2)} P^{(2)}(\Omega, k_x, z) e^{jk_z(\Omega).z} \qquad 9(b)$$

Similar to Eq. 9(b), we obtain the corresponding 1-D differential equation in $z$ for the optical field in Eq. (10).

$$\frac{\partial A_{op}(\omega, k_x, z)}{\partial z} = -\frac{jk_{x0}(\omega)k_x}{k_{z0}(\omega)} A_{op}(\omega, k_x, z) + \frac{jk_x^2}{2k_{z0}(\omega)} A_{op}(\omega, k_x, z)$$
$$-\frac{j\omega^2}{2k_{z0}(\omega)\varepsilon_0 c^2} P_{op}(\omega, k_x + k_{x0}, z) e^{jk_{z0}(\omega)z} \qquad (10)$$

In Eq. (10), the first term on the right hand side represents the oblique propagation of the optical field. The second term corresponds to the quadratic phase associated with a Gaussian beam and the final term is the spatial Fourier transform of the nonlinear polarization in Eq. (6). Equations 9(b) and (10) form a system of coupled 1-D differential equations for each value of $k_x$. Equations (5), (6) and 9(b) and (10) are thus solved progressively up to $z = h\sin\alpha$.

*4.4 Calculating Transmission and Propagation of THz radiation*

To calculate the transmitted THz field, we use standard Fresnel reflection coefficients as a function of the transverse-momentum $k_x$. For THz field polarization perpendicular to the plane of Fig. 1, the Fresnel reflection coefficients are presented in Eq. (11).

$$T(k_x) = 2\sqrt{\frac{\Omega^2 n(\Omega)^2}{c^2} - k_x^2} \left(\sqrt{\frac{\Omega^2 n(\Omega)^2}{c^2} - k_x^2} + \sqrt{\frac{\Omega^2}{c^2} - k_x^2}\right)^{-1} \qquad (11)$$

In [28], special reflection coefficient formulae for the case when the group velocities of the optical and THz radiation are disparate were provided. However, close to phase-matching this is less important. The THz field at a distance $z_d$ away from the exit surface is calculated without paraxial approximations as shown in Eq. (12).

$$E(\Omega, x, h\sin\alpha + z_d) = \mathbf{F}_x^{-1}\left(A(\Omega, k_x, h\sin\alpha)T(k_x)e^{-j\left(k_0^2(\Omega)-k_x^2\right)^{1/2}z_d}\right) \quad (12)$$

Thus using Eqs. (1)-(12), we can model an arbitrary TPF setup for THz generation by OR. Equations 9 and 10 are effectively a 1-D system of coupled equations and can be solved in parallel for various $k_x$ and $\omega$ and $\Omega$. Numerical integration was performed using a 4$^{th}$ order Runge-Kutta method. The evaluation of Fourier transforms was accelerated by GPU parallelization.

## 5. Validation of the model

| Optical Element | ABCDEF Matrix |
|---|---|
| Diffraction grating | $A = \dfrac{\cos\theta_d}{\cos\theta_i}$, $B=0$, $C=0$, $D = \dfrac{\cos\theta_i}{\cos\theta_d}$, $E=0$, $F(\omega) = F_0 + F_1 + F_2$ |
| | $F_0 = \dfrac{-2\pi c(\omega - \omega_0)}{p\omega_0^2 \cos\theta_d}$ |
| | $F_1 = F_0\left(-\dfrac{(\omega-\omega_0)}{\omega_0} + \dfrac{1}{2}\tan\theta_d F_0\right)$, $F_2 = F_0\left(\left(\dfrac{\omega-\omega_0}{\omega_0}\right)^2 + \dfrac{1}{2}\tan\theta_d F_1 + \dfrac{F_0^2}{6}\right)$ |
| | $\sin\theta_i + \sin\theta_d = \dfrac{2\pi c}{p\omega}$ |
| Propagation in a medium | $A=1, B=L/n(\omega), C=0, D=1, E=0, F=0$ |
| Lens | $A=1, B=0, C=-1/f, D=1$ |

Table 1: ABCDEF ray pulse matrices for various optical components are presented. The diffraction grating has terms with upto 4$^{th}$ order dependence on frequency which accounts for group velocity dispersion due to angular dispersion and higher order dispersive terms.

In this section, the developed model is validated against analytic theory and experiments. Simulations assume the setup shown in Fig. 1(a). The ABCDEF matrices for the various optical components are presented in Table.1. The $F$ term for the diffraction grating contains dispersive terms upto the 4$^{th}$ order in frequency and accounts for GVD-AD and higher order dispersion terms. The full-width at half maximum (FWHM) pulse duration was assumed to be 0.5 ps with a fluence of 20 mJ/cm$^2$ and a peak intensity of 40 GW/cm$^2$. The effective second order susceptibility was assumed to be $\chi_{eff}^{(2)}(x,z) = 360$ pm/V [29] and the intensity dependent refractive index coefficient was $n_2=10^{-15}$ cm$^2$/W [30]. The optical beam with input $e^{-2}$ radius of $w_{in} = 2.5$ mm and is incident at $h = 1.5$ mm from the apex of the crystal. The focal length of the lens was 23 cm. The refractive indices and Raman gain lineshape are taken from [31] and [32] respectively. We calculate the conversion efficiency as a function of the grating incidence angle ($\theta_i$), grating to lens ($s_1$) and lens to crystal ($s_2$) distances. The values of $\theta_i$, $s_1$ and $s_2$ obtained for maximum conversion efficiency are compared to analytic calculations [21].

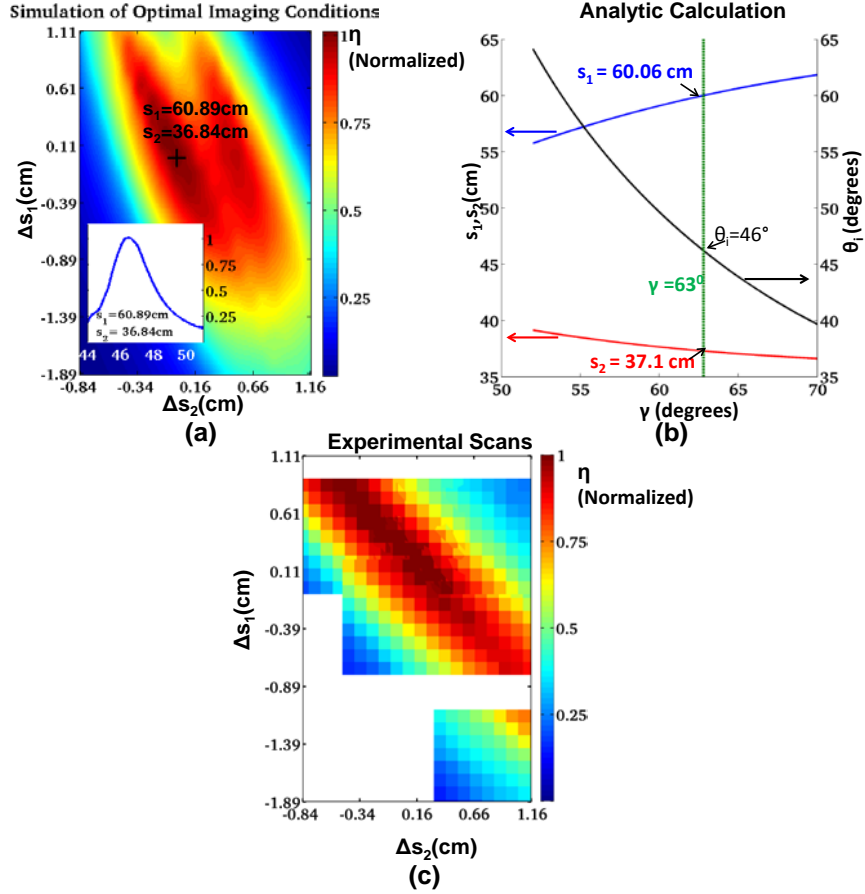

**Fig.4**(a): Simulation of conversion efficiency as a function of imaging conditions. The surface plot shows the conversion efficiency versus displacements from optimal imaging distances $\Delta s_1$ and $\Delta s_2$. $s_1$, $s_2$ are the lens-to-grating and lens-to-crystal distances respectively. The inset shows conversion efficiency versus incidence angle to diffraction grating. As $s_1,s_2$ are varied, there is variation of the pulse-front-tilt angle which leads to a change in conversion efficiency. Careful optimization of the experimental setup is required to identify the optimal conversion efficiency point. (b) Theoretical calculations of optimal imaging conditions for various pulse-front-tilt angles based on analytic theory from [21]. For a pulse-front-tilt angle of 63°, the imaging conditions are in close agreement with the simulation results, validating the accuracy of the presented model. (c) Experimental scans of conversion efficiency vs displacements $\Delta s_1$ and $\Delta s_2$ agree well with the simulations in Fig. 4(a).

The simulation results are plotted in Fig. 4(a). A maximum conversion efficiency is obtained at $s_1 = 60.89$ cm, $s_2 = 36.84$ cm and $\theta_i = 46.5°$. It can be seen that these values of $s_1$, $s_2$ satisfy the imaging condition for the setup, i.e. $s_1^{-1} + s_2^{-1} = f^{-1}$. The optimal imaging condition is determined by the magnification required to produce the optimal pulse-front-tilt angle inside the crystal. Analytic calculations were developed in [21] to supply optimal imaging conditions and are presented in Fig. 4(b). The blue and red curves in Fig. 4(b) correspond to the values of $s_1$ and $s_2$ for various pulse-front-tilt angle values $\gamma$ and are plotted along y-axis on the left. The grating incidence angle $\theta_i$ as a function of $\gamma$ is given by the black curve plotted along the y-axis on the right. We know that for pumping at 1030 nm, the optimum pulse-front-tilt angle $\gamma = \cos^{-1}(n_g(\lambda_0)n(\Omega)^{-1}) \sim 63°$. For this value of $\gamma$, it is seen that the corresponding imaging conditions closely match the simulation results, i.e. $s_1 = 60.06$ cm, $s_2 = 37.1$ cm and $\theta_i = 46°$. Since the simulation is *ab-initio*, i.e it does not make any prior assumptions about the optimal imaging conditions, this agreement is a strong validation of the developed model. As $s_1$ or $s_2$ are varied, the direction $x'_{out}(\omega)$ at which each optical component at $\omega$ emerges from the setup changes which affects phase-matching via Eq. (5). Thus, the presented formalism can map the performance of the system directly to experimental conditions. In Fig. 4(a), we see how small deviations in imaging conditions can lead to sizeable degradation of conversion efficiency. For example, $\Delta s_2 \sim 1$ mm, leads to drop in conversion efficiency by about 40%. In Fig. 4(c), we show experimental scans of conversion efficiency for similar parameters. The white spaces in the figure indicate regions where data was not collected. For similar displacements $\Delta s_1$ and $\Delta s_2$ from the optimum values, the conversion efficiency reduction agrees well with the calculations in Fig. 4(a). The slight difference in the tilt of the ellipse in Fig. 4(c) can be attributed to a different grating incidence angle from that presented in Fig. 4(a).

Further verification of the model is provided by comparisons to experiments [14]. For the optical pump conditions described for Fig. 4(a), a conversion efficiency of 0.8% ($w_{in}$= 3.5 mm $h$ =2.2 mm) which is in reasonable agreement with the experimentally reported value of 1.15%. The larger number may be partially owed to uncertainties in THz absorption coefficients below 0.9 THz [31]. When the conversion efficiency is 0.8%, the absorption coefficient below 0.9 THz was ~ 10 cm$^{-1}$. When this was adjusted to 5 cm$^{-1}$, the resulting conversion efficiency was 0.9%, which is closer to the experimental result.

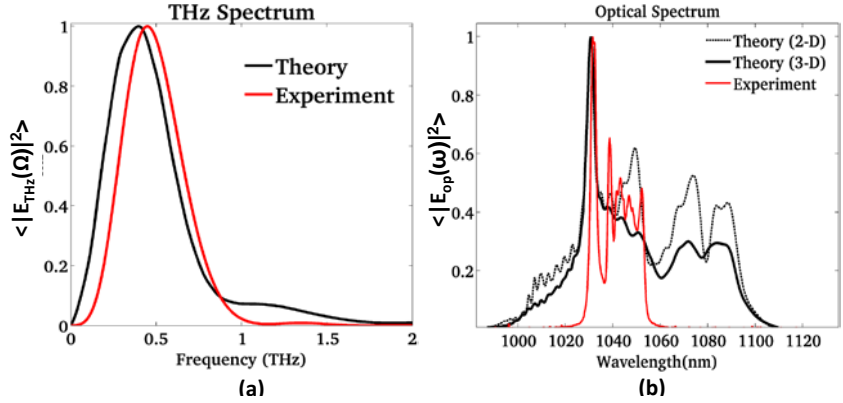

Fig. 5 (a) The experimental and theoretically calculated THz spectra are presented. The theoretical calculation is spatially averaged over $x$ and is centred at 0.45 THz, in close agreement with experiments [14] (b) The experimental output optical spectrum (red) is presented along with calculations. The theoretical calculation averaged over a single transverse spatial dimension $x$ (black-dotted) are broadened significantly more than the experiments. However, if the spatial averaging is performed over both transverse spatial dimensions $x$ and $y$ by simulating numerous 2-D slices, the output spectrum matches experiments more closely. The disparity may also be partially explained by the possibility of incomplete collection of extreme optical frequency components with large divergence.

In Fig. 5(a), the experimentally obtained THz spectrum is compared to theoretical calculations. The calculation presented is the spatially averaged spectrum $\left\langle \left| E_{THz}(\Omega) \right|^2 \right\rangle$ $= \int_{-\infty}^{\infty} \left| E_{THz}(\Omega, x, h\sin\alpha) \right|^2 dx$ at the output facet of the crystal. The theoretical and experimental spectra are in agreement and peak at ~0.45 THz, in line with expectations from prior models [18], [21]. In Fig. 5(b), the experimentally reported optical spectrum is compared to theoretical calculations. The black dotted line represents spatial averaging of the optical spectrum over a single transverse dimension, i.e

$\left\langle \left| E_{op}(\omega) \right|^2 \right\rangle_{2-D} = \int_{-\infty}^{\infty} \left| E_{op}(\omega, x, h\sin\alpha) \right|^2 dx$. The theoretically obtained optical spectrum shows significant broadening and red-shift similar to the experiments. However the theoretically obtained spectrum is broadened to a larger extent than the experimental spectrum. We simulated several 2-D slices, each with different optical pump intensity, and averaged the optical spectrum over two spatial dimensions $\left\langle \left| E_{op}(\omega) \right|^2 \right\rangle_{3-D} = \int_{-\infty}^{\infty} \int_{-\infty}^{\infty} \left| E_{op}(\omega, x, y, h\sin\alpha) \right|^2 dxdy$. This is shown in the solid black curve and is in better agreement with experiments. The reason for this is that for lower optical intensities, the extent of spectral broadening will be less as the nonlinear polarization term in Eqs.(5)-(6) would be smaller and therefore the spatial average shows less red-shift and spectral broadening. Another reason for the disparity between experiments and theory could be attributed to uncertainties in the measurement of the optical spectrum. Extremities of the frequency spectrum may not have been collected due to their larger divergence.

## 6. Optimizing conversion efficiency

*6.1 Discussion of effective length in two dimensions*

It is useful to understand what the effective propagation length $L_{eff}$ in a 2-D geometry is. In general, absorption and dispersion determine the optimal value of $L_{eff}$. A longer $L_{eff}$ leads to more absorption and dispersion. A shorter effective length would mean less of both but would also translate to lesser THz generation. Therefore, there must exist an optimum value where the amount of THz generation is sufficiently large but absorption and dispersion are small.

In a 2-D non-collinear geometry, two parameters influence the extent of absorption and dispersion. These include the beam radius $w_{in}$ (or $w_{out}$) and beam position $h$ of the optical pump beam (with respect to the apex of the crystal, see Fig. 1 (b) for definitions). Therefore, the effective length parameter maps to both $h$ and $w_{in}$, i.e. $L_{eff} \to g(h, w)$ where $g$ is some function. Since THz is generated only in the region where there is optical fluence, (i) a larger value of $h$ would mean that the THz would propagate over a longer absorptive region (i.e regions where this no optical fluence, e.g. see Figs. 6(a) vs. 6(b)). Mathematically, the regions without optical fluence correspond to regions where the nonlinear polarization term $P_{THz}(\Omega, k_x, z) = 0$ in Eq. 9(b). (ii) For larger $h$, the optical beam would propagate longer distances which would mean it would suffer greater spectral broadening due to cascading effects and SPM. This would in turn make dispersive effects more acute. Thus a larger $h$ increases both dispersion and absorption. At the same time, a small $h$ also leads to lesser amount of THz being generated. (iii) A smaller value of beam radius $w_{in}$ would lead to a larger region without optical fluence and consequently a larger amount of absorption. (iv) A larger value of $w_{in}$ would mean that parts of the optical pump beam propagate a longer distance (since different sections of the beam propagate different distances), which would lead to a greater amount of cascading and dispersive effects. Therefore, it is reasonable to expect an optimal value of $w_{in}$ and $h$ for a given set of optical pump conditions. Increasing the intensity or initial bandwidth of the optical pulse will lead to more rapid spectral broadening (w.r.t length), which would require a readjustment of $h$ and $w_{in}$.

*6.2 Implications on energy scaling*

These effects are illustrated in Figs. 6(a)-(c). The fluence, bandwidth, pulsewidth and material parameters are the same as that used for Figs. (4). In Fig. 6(a), a beam with $w_{in}$ = 2.5 mm, is incident at $h$ = 1.5 mm from the crystal apex. It can be seen how the THz and optical beams have good overlap which reduces absorptive effects and results in a relatively high conversion efficiency of 0.7%. In Fig. 6(b), the same beam is displaced further down the crystal. One sees that increased THz absorption as delineated in Fig. 6(b), causes the conversion efficiency to drop to 0.3%. In Fig. 6 (c), a larger beam size with $w_{in}$ = 10 mm is used at $h$=5 mm. We see that only a small portion of the optical pump beam cross-section produces THz radiation, resulting in a conversion efficiency of 0.5%. This is because, after initial THz generation, subsequent parts of the beam are spectrally broadened due to cascading effects to an extent that prevents further THz generation in the presence of GVD-AD and GVD-MD. This has an important implication on the scaling of these systems to large pump energies. In Fig. 6(d), we plot the maximum conversion efficiencies for various values of $w_{in}$ while keeping the peak optical pump intensity constant. The top *x*-axis depicts the corresponding optimal values of *h*. Note that the size of the beam at the input crystal face is $w_{out} \sim 0.6 w_{in}$. In Fig. 6(d), for $w_{in}$ > 3.5 mm, there is a drastic drop in the maximum achievable conversion efficiency. There is an initial increase in the maximum achievable conversion efficiency for $w_{in}$ < 3.5mm, because of reduced absorption due to the increase in beam size (See section 6.1 for explanation). In Fig. 6(d), it is seen that optimal values of *h* increase with larger $w_{in}$ and that they are present relatively close to the apex of the crystal. The degradation of conversion efficiency can be circumvented by using an elliptical pump beam with its major axis perpendicular to the plane of tilting (i.e out of the plane of the paper).

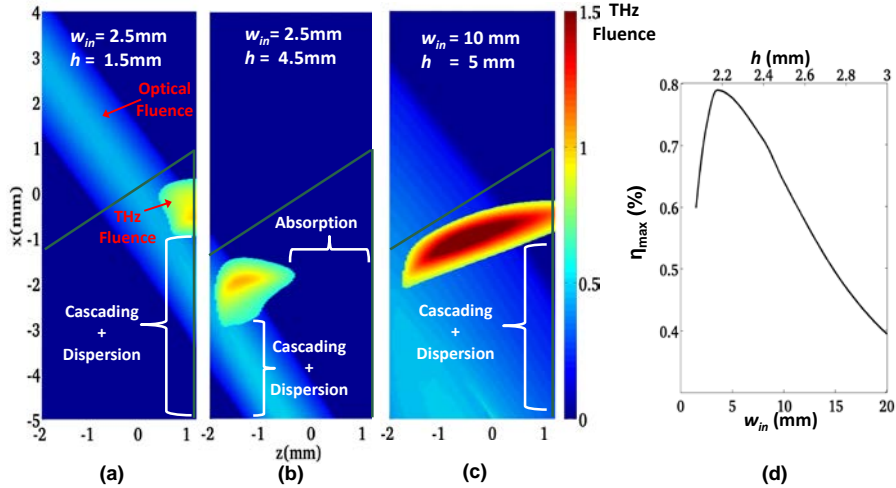

Fig. 6(a). Effective length in 2-D: The beam position $h$ and beam radius $w_{in}$ affect the amount of absorption and dispersion and therefore are mapped to the effective length in 2-D. Absorption is proportional to $h$ since there would be a greater region of space without optical fluence for larger $h$, whereas $w_{in}$ is proportional to the area containing optical fluence and is therefore proportional to the amount of absorption. An increase in $h$ or $w_{in}$ increases the effective propagation distance of the optical beam and therefore dispersive effects due to spectral broadening caused by cascading effects. Simultaneously, a very small $h$ or $w_{in}$ results in lesser THz generation. Therefore, there is an optimal $h$ and $w_{in}$ for each optical pump condition. (a) For $h$=1.5 mm and $w_{in}$=2,5 mm there is minimal absorption and conversion efficiency is 0.7% (b) for $h$=4.5 mm absorption increases and conversion efficiency drops to 0.3 % (c) for large $w_{in}$=10 mm, only small portions of the beam are involved in THz generation due to disruption of phase-matching by enhanced dispersive effects in the presence of cascading effects , leading to an overall drop in conversion efficiency to 0.4 %. This has important implications on the scaling of THz energies, by merely scaling beam size. (d) For $w_{in}$ < 3.5 mm, absorptive effects dominate and conversion efficiency increases with beam size. The maximum achievable conversion efficiency drops beyond $w_{in}$=3.5 mm, due to enhanced dispersive effects caused by cascading effects. An elliptically shaped beam is thus preferred for very large conversion efficiencies.

### 6.3 Effects of pump intensity

In Fig. 7(a), we experimentally determine the conversion efficiency as a function of fluence for two different cases. In these experiments, a pulse with duration of 0.5ps was stretched to 1.39 ps. In the black curve, the setup is optimized to yield maximum conversion efficiency at the highest peak intensity by adjusting $h$. The fluence is then progressively decreased. In the red curve, the conversion efficiency is optimized for the lowest fluence and then progressively increased. The curves show a hysteresis with a cross-over at a fluence of 22 mJ/cm$^2$. In Fig. 7(b), we simulate these experiments using the developed model. The simulated results show qualitative and quantitative agreement with the experiments. For lower fluences, a larger value of $h$ is required to optimize the conversion efficiency. This is because, the smaller peak intensity in this case leads to a slower rate of spectral broadening of the optical pump, thereby causing dispersive effects to be 'delayed' in their appearance. This leads to a longer effective length or larger $h$ for optimum efficiency. Note that the optimal values of $h$ are larger compared to those depicted in Fig. 6(d) in Section 6.2 because of the smaller peak intensity of the stretched pulse. Thus we see that the conditions for the optimal conversion efficiency are highly sensitive to optical pump conditions. This could be one reason why, various experiments report very different saturation curves.

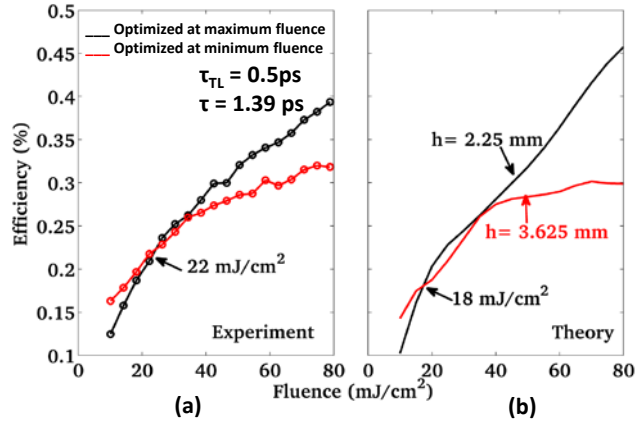

Fig. 7(a): Experimentally obtained conversion efficiency saturation curves optimized for different pump intensities. The black curve is optimized for the maximum intensity while the red curve is optimized for the smallest intensity. The optimal experimental conditions are different for different intensities as seen in the hysteresis of the curve (b) Theoretical calculations of conversion efficiency saturation curves for experimental parameters in (a). Good quantitative agreement between experiments and theory is seen. When the intensity is lower, the optimal efficiency occurs at a larger value of $h$. This is because cascading effects occur at a slower rate and enable a longer effective interaction length. The optimal values of $h$ are larger than that in Fig. 6(d) due to the use of a stretched pulse in the experiments.

*6.4 Trade-offs of optimizing conversion efficiency*

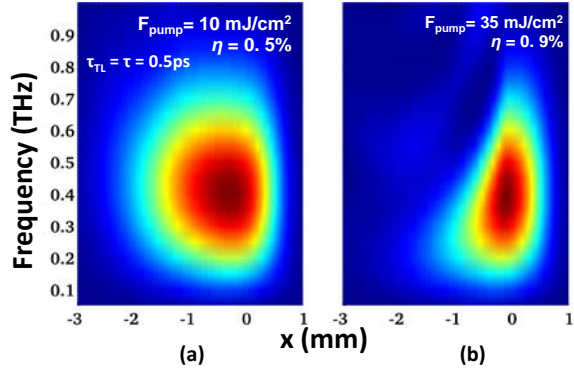

Fig. 8 (a). When the fluence is 10 mJ/cm$^2$, the conversion efficiency is 0.5%. The THz spectrum as a function of transverse co-ordinate *x*, is relatively uniform with all points having a broadband THz spectrum centred at ~0.45 THz. (b) As the fluence is increased to 35 mJ/cm$^2$, the conversion efficiency increases to 0.9% but the THz beam now contains a large spatial chirp and has an effectively reduced spot size. As the optical beam propagates to more negative values of *x*, it has been significantly broadened spectrally. Along with dispersive effects, this inhibits further coherent growth of THz radiation. Absorptive effects then dominate, leaving only the lower frequency THz components with smaller absorption intact.

Finally, we highlight some trade-offs of optimizing only conversion efficiency in Figs. 8(a) and 8(b). Here, the THz spectrum as a function of transverse spatial co-ordinate (*x*) are shown for two different fluences. In Fig. 8(a), the optical fluence is 10 *mJ/cm$^2$* which results in a conversion efficiency of 0.5%. The THz spectrum is virtually identical across the beam cross-section (*x* co-ordinate*)* as seen in Fig. 8(a). In Fig. 8(b), a conversion efficiency of 0.9% is achieved with a fluence of 35 *mJ/cm$^2$*. However, the THz spectrum is spatially chirped across the beam-cross section and has a effectively smaller spot-size. As the optical pump beam generates THz, it suffers spectral broadening due to cascading effects. Phase mismatch is accentuated in the presence of this larger spectral bandwidth by GVD-AD and GVD-MD, which causes the coherent growth of THz to cease at the more negative values of *x* (See Fig. 2(a) : optical beam propagates towards negative *x* values). In the absence of coherent growth, THz absorption dominates and the THz spectrum red-shifts (since absorption is less for lower THz frequencies). Thus, while conversion efficiency is increased, a spatial chirp is introduced in the THz beam profile along with a reduction in the THz beam spot-size which may not be suitable for certain applications.

*7. Conclusion and Future Outlook*

In conclusion, a new approach to modelling THz generation via optical rectification (OR) using tilted-pulse-fronts (TPF's) was presented and discussed. The approach was formulated to consider (i) spatio-temporal distortions of the optical pump pulse, (ii) coupled non-linear interaction of the THz and optical fields in 2-D as well as (iii) self-phase-modulation and (iv) stimulated Raman scattering. The formulation was done in a way to circumvent challenging numerical issues. It was validated by comparisons to experiments and analytic calculations, with good quantitative agreement in both cases. We described the physics of OR using TPF's. In particular, we discussed the problem from transverse-momentum and time domains. It was seen that the large spectral broadening which accompanies THz generation also leads to broadening in transverse-momentum since each frequency component in an angular dispersed beam has a well-defined transverse-momentum value. Thus, the optical pump as an increased frequency and transverse-momentum spread. Group velocity dispersion due to angular and material dispersion then causes a spatio-temporal break-up of the optical pump pulse which inhibits further coherent build-up of THz radiation. It is seen that THz conversion efficiency reduces for very large beam sizes. This suggests the use of optical pump beams that are elliptically shaped for high energy pumping. Guidelines to optimize the setup are provided. Imaging errors were shown to be critical and careful alignment is required to optimize efficiency. It is seen that optimal setup conditions are different for different pump conditions. Finally, we show how optimizing the conversion efficiency could lead to other trade-offs such as a deterioration of the spatial THz beam profile. This work provides an overview for optimizing such sources for various applications.